\documentstyle[aps,epsf,amssymb,graphicx]{revtex}
\begin{document}
\tightenlines
\draft
\title{A strong dissipative inflationary model}
\author{Mauricio Bellini\footnote{E-mail address: mbellini@mdp.edu.ar}}
\address{Consejo Nacional de Investigaciones Cient\'{\i}ficas y
T\'ecnicas (CONICET)\\
and\\
Departamento de F\'{\i}sica,
Facultad de Ciencias Exactas y Naturales,
Universidad Nacional de Mar del Plata, \\
Funes 3350, (7600) Mar del Plata, Buenos Aires, Argentina.}
\maketitle
\begin{abstract}
The dynamics of a single scalar field which describes a second order
phase transition is considered taking into account the thermal
effects in the high temperature limit. I conclude that thermal
effects play a very important role during inflation and do not must
to be ignored in models for the early universe. At the end of inflation
thermal fluctuations are more important than quantum fluctuations, due
to the fact that $T>H_c$.
\end{abstract}
\vskip .2cm                             
\noindent
Pacs numbers: 98.80.Cq\\
\vskip 1cm

\section{Introduction}

An inflationary phase eliminates the major problems in cosmology, namely
flatness, the horizon problem, homogeneity and the numerical density
of monopoles\cite{1a,2a,3a,4a,linde90,starobinsky}.
However, the main problem is how to attach
the observed universe to the end of the inflationary epoch.
In the standard (isentropic) inflationary scenarios, the radiation
energy density $\rho_r$ scales with the inverse fourth power
of the scale factor, becoming quickly negligible. In such case, a
short time reheating period terminates the inflationary period
and initiates the radiation dominated epoch. On the other hand,
the only condition required by general relativity for inflation
is that $\rho_r < \rho_v$. Inflation in presence of non-negligible
radiation is characterized by nonisentropic expansion, and thermal
seeds of density fluctuations.
Within a realistic quantum field theory model, Hosoya and Sakagami\cite{1}
initially formulated dissipative dynamics. In their formulation, the
near-themal-equilibrium dynamics is expressed through an expansion
involving equilibrium correlation functions. Subsequently Morikawa\cite{mor}
formulated the same problem in terms of an elegant real time finite temperature
field theory formalism. These works were developed further by Lawrie\cite{law}
and Gleiser and Ramos\cite{gr}. For models like warm inflation\cite{wi} the
overdamped regime of the inflaton is the one of interest. A realization of
overdamped motion within quantum field theory model based on this formulation
was obtained in\cite{2}. A more recently warm inflationary model inspired
in string theories was developed in\cite{bgrr}.

In this paper I study the equation of motion for the expectation
value of the inflaton
field $\varphi$ on a thermodynamical average for a symmetric massive
scalar field with self-interaction.
Unfortunately no reliable theory of a system far from equilibrium has yet
been available. We have to content ourselves to work in the
hydrodynamical regime, i.e., slight-off-equilibrium state following
some ideas of Hosoya and Sakagami\cite{1}, who obtained this equation
which also includes the effect of the temperature increase by friction
during inflation.
For this propose I will consider the equation of motion $\Box \varphi +
V'(\varphi) = 0$ on a background
spatially flat Friedmann-Robertson-Walker (FRW)
$ds^2 = -dt^2 + a^2 dx^2$
metric on a thermodynamical average by means of the semiclassical
approach $\varphi(\vec x,t) = \phi_c(t) + \phi(\vec x,t)$.
Here, $\phi_c(t)$
is given by $\left< \varphi(\vec x,t) \right>={\rm tr}(\rho\varphi)$
for a suitable nonequilibrium statistical operator $\rho$
and $\phi(\vec x,t)$ is the fluctuation field around the
thermodynamical average value $\phi_c$. Furthermore,
the operator $\phi$ has a zero thermodynamical average $\left<\phi\right>=0$.
In the Heisenberg representation the operator $\rho$ is time independent
(i.e., ${d\rho \over dt} = 0$). The classical scalar field $\phi_c$
starts at a
point a little bit away from the top and then rolls down along
the long slope to the minumum of the effective potential.
If the friction force is strong enough, there will be
no oscillation and $\phi_c$  simply slowly rolls down to the bottom.

\section{The effective equation of motion}

We consider the thermodynamical average of the field equation
$\Box \varphi + V'(\varphi) =0$
\begin{equation}\label{1}
\left< \Box \varphi + V'(\varphi) \right> =0,
\end{equation}
where
\begin{displaymath}
\left<\Box \varphi\right>= (\ddot\phi_c +\ddot\phi) - {1\over a^2}
\nabla^2\phi + 3 \left< H(\phi_c+\phi)  (\dot\phi_c+\dot\phi)\right>.
\end{displaymath}
For simplicity, I will consider $\left<H(\phi_c+\phi) \right>
\simeq H_c(\phi_c)$
where 
$H_c(\phi_c) = \dot a/ a$ is the Hubble parameter and $a$ is the
scale factor of the universe. The dot and prime denote the derivatives
with respect to time and the field, respectively.
On the other hand, the thermodynamical average $\left<V'(\varphi)\right>$
can be expanded as
\begin{equation}\label{2}
\left<V'(\phi_c + \phi)\right> = V'(\phi_c) + V''(\phi_c) \left<\phi\right>
+ \frac{V'''(\phi_c)}{2!} \left<\phi^2\right> + ... .
\end{equation}
Hence, the equation of motion for the classical field being
given by
\begin{equation}\label{3}
\ddot\phi_c + 3 H_c \dot\phi_c  + V'(\phi_c) + \frac{V'''(\phi_c)}{2!}
\left<\phi^2\right> + ... =0.
\end{equation}
The thermodynamical average of the squared fluctuations are
\begin{equation}\label{4}
\left<\phi^2\right> = {\Large\int} \frac{d^3p}{(2\pi)^3} \frac{1}{\omega(p)}
n(p),
\end{equation}
where $\omega^2(p)=[\vec p^2 + V''(\phi_c)]$ and $n(p)$ is the
bosons distribution given by $n(p) = n_0(p) + \delta n(p)$ in the
first order approximation. The local equilibrium
distribution $n_0(p)$ is given by
\begin{equation}\label{5}
n_0(p) = \frac{1}{e^{\beta\omega(p)}-1},
\end{equation}
where $\beta=1/T$ is the inverse of the temperature $T$.
Furthermore, $\delta n(p)$ is a small deviation of $n_0$.
If $\tau$ is the relaxation time of the fluctuations, in the relaxation-time
approximation $\delta n = -\tau \partial_t n_0$
\begin{equation}\label{6}
\delta n = \tau n_0 \left[ \frac{\beta}{2\omega(p)}
V'''(\phi_c) \dot\phi_c + \dot\beta \omega\right],
\end{equation}
where was used $\dot\omega = {V'''(\phi_c) \dot\phi_c \over
2\omega}$ and $d_t \equiv \dot\phi_c d_{\phi_c}$.
Furthermore, $\tau$ is the relaxation time for the fluctuations.

\subsection{Calculation of $\left< \phi^2\right>$}

From eqs. (\ref{4}), (\ref{5}) and (\ref{6}) the thermodynamical average
of the squared fluctuations $\phi$ are
\begin{eqnarray}
&& \left<\phi^2\right> = \left<\phi^2\right>_0 + \left[
\dot\phi_c V'''(\phi_c)
{\Large\int} \frac{d^3p}{(2\pi)^3} \frac{\tau n_0 (1+n_0) \beta}{2\omega^2(p)}
\right.\nonumber \\
&& +\left.{\Large\int} \frac{d^3p}{(2\pi)^3}
\tau n_0 (1+n_0) \dot\beta\right], \label{7}
\end{eqnarray}
where the first term in eq. (\ref{7}) gives the fluctuations
in thermal equilibrium
\begin{equation}\label{8}
\left< \phi^2\right>_0 = {\Large\int} \frac{d^3p}{(2\pi)^3} \frac{n_0(p)}{
\omega(p)},
\end{equation}
and the remaining terms in eq. (\ref{7}) describe the contribution
out of the thermal equilibrium. Replacing eq. (\ref{5}) in the equation
(\ref{8}) and integrating, we obtain the following result in the
high temperature limit (i.e., for $\beta m \ll 1$):
\begin{equation}
\left<\phi^2\right>_0 \simeq \frac{T^2}{12} \left(1 - \frac{4m}{\pi^2 T}\right).
\end{equation}
On the other hand, using the fact that $\tau^{-1}
\simeq {\lambda^4 \over 3. 2^9 \pi} {T^2 \over \omega}$\cite{1} in the
high temperature limit, the first integrate in eq. (\ref{7}) gives
\begin{equation}\label{9}
{\Large\int} \frac{d^3p}{(2\pi)^3}
\frac{\tau n_0 (1+n_0) \beta}{\omega^2(p)} \beta \simeq
\frac{3. 2^8}{\pi T} {\rm ln}\left(T/m\right).
\end{equation}
The second integrate in eq. (\ref{7}) results
\begin{equation}\label{10}
{\Large\int} \frac{d^3p}{(2\pi)^3} \tau n_0 (1+n_0) \dot\beta \simeq
- \frac{3. 2^7}{\pi} \dot T .
\end{equation}
We can obtain $\dot T$ from the equation which describes the evolution
for $\rho_r={\pi^2 \over 30} g_{eff} \  T^4$
\begin{equation}\label{11}
\dot\rho_r + 4 H_c \rho_r - \delta =0.
\end{equation}
If there were no dissipation, then $\delta =0$ and the radiation
component would be rapidly red-shifted away
as $\rho_r \sim e^{-int 4 H_c dt}$. However with dissipation, radiation
is being produced continuously from conversion of scalar field
vacuum energy.
In our case
$\delta = \eta \dot\phi_c^2$ describes the interaction term due to
the decay of the inflaton field without consider the temporal
evolution of the temperature due to the expansion of the universe.
We are interested in the study of models where $\dot\rho_r >0$
(Fresh inflation)\cite{prd2001}. Inflationary models where the radiation
energy density remains almost constant ($\dot\rho_r \simeq 0$) are
named Warm Inflation\cite{wi}.
I will consider the potential
\begin{equation}\label{12}
V(\phi_c) = \frac{m^2}{2} \phi_c^2 + \frac{\lambda^2}{24} \phi_c^4 + \Lambda,
\end{equation}
where $\Lambda $ is the cosmological constant.
For this potential we obtain the following dissipative parameter:
\begin{eqnarray}
\eta & = & \beta \left[ \frac{V'''(\phi_c)}{2}\right]^2 {\Large\int}
\frac{d^3 p}{(2\pi)^3} \tau \frac{n_0}{\omega^2} (1+n_0) \nonumber \\
& \simeq & \frac{3.2^6}{\pi} \frac{\phi_c^2}{T} {\rm ln}
\left(\frac{T}{m}\right).
\label{13}
\end{eqnarray}
Hence, from eqs. (\ref{11}) and (\ref{13}), we obtain the temporal evolution
for $T$ in eq. (\ref{10})
\begin{equation}\label{14}
\dot T = \frac{45. 2^5}{\pi^3 g_{eff}} \frac{{\rm ln}(T/m)}{T^4} \dot\phi_c^2
\phi_c^2 - H_c T.
\end{equation}
Note that eq. (\ref{14}) describes the evolution of the temperature
taking into account dissipation and expasion effects.
The first term in eq. (\ref{14}) comes from dissipative effects and
the second one takes into account the expansion of the universe.
For this reason, for $\eta =0$ we obtain the red-shifted temperature
$T \sim e^{-\int H_c dt}$ (which would be $T \sim e^{-H_0 t}$, with
$H_0 = {\rm const.}$, in a de Sitter expansion).
In our case dissipative effects are more important than the expansion
of the universe (i.e., $\eta > H_c$), so that the temperature
increases: $\dot T >0$.

Finally, the squared fluctuations $\left< \phi^2 \right>$ in the high
temperature limit, are
\begin{eqnarray}
&& \left.\left< \phi^2(\vec x,t)\right>\right|_{T \gg m}
\simeq \frac{T^2}{12} \left[1-
\frac{4 m}{\pi^2 T} \right] \nonumber \\
&& +
\frac{3. 2^7}{\pi} \left\{ \frac{{\rm ln}(T/m)}{T}
\left[\dot\phi_c\phi_c - \frac{45. 2^5}{\pi^3 g_{eff} T^3} \phi^2_c \dot\phi_c^2
\right] + H_c T\right\}. \label{15}
\end{eqnarray}

\subsection{Inflationary dynamics}

Once obtained the equation that describes the temporal evolution for
the temperature [see eq. (\ref{14})] and the squared field fluctuations
$\left<\phi^2\right>$ we can write the effective equation of motion
for the background field $\phi_c$
\begin{equation}\label{16}
\ddot\phi_c + \left[ 3 H_c + \eta_{eff}\right] \dot\phi_c +
V'_{eff}(\phi_c,T) =0,
\end{equation}
with $\eta_{eff}$ and $V'_{eff}$ given by
\begin{eqnarray}
\eta_{eff} & = & \eta - \frac{135. 2^{11}}{\pi^4 g_{eff}}
\frac{{\rm ln}(T/m)}{T^4} \phi_c^3 \dot\phi_c,  \label{17} \\
V'_{eff}(\phi_c,T) & = & \left[ m^2 + \frac{\lambda^2 T^2}{24}
\left(1- \frac{4 m}{\pi^2 T}
\right)\right] \phi_c \nonumber \\
& + & \frac{\lambda}{6} \phi^3_c +
\frac{3. 2^6}{\pi} H_c \phi_c T. \label{18}
\end{eqnarray}
The additional term in $\eta_{eff}$ comes from the first term in eq.
(\ref{14}) which is originated by the interaction of the inflaton field
with the background thermal bath. In other
words $\eta_{eff}$ is given by the $\eta$-self-interaction of the inflaton
field plus the additional dissipative contribution
given by the fact that the fields produced by self-interaction interact with
the fields of the background thermal bath. Futhermore, the remaining
term in eq. (\ref{14}) is conservative and can be added in the
$\phi_c$-derivative of the
effective potential $V_{eff}(\phi_c,T)$[see eq. (\ref{18})].
This term is originated by the
expansion of the universe and its origin is purely cosmological.

It is very important to take into account that the equation (\ref{16})
is valid only if the hydrodynamical condition is satisfied:
\begin{displaymath}
\tau \ll \left|\frac{a}{\dot a}\right|,
\left|\frac{\phi_c}{\dot\phi_c}\right|.
\end{displaymath}
This means that the relaxation time of the fluctuations must to be
very small with respect to the inverse of the Hubble parameter and
the characteristic time of change for the background field $\phi_c$.
Furthermore, one has to be careful about the region in which eq. (\ref{16})
is applicable. This equation is only valid in the limit of high temperatures
and the condition $T/m \gg 1$ must to be fulfilled.

Finally, the Friedmann equation $ \left<H^2(\varphi)\right> =
8\pi G/3\left[\left<\rho_{\phi} +
\rho_r\right>\right]\simeq 8\pi G/3\left[\left<\dot\varphi^2/2 \right>
+ \left< V(\varphi)\right>\right]$ for an inflating universe with
potential (\ref{12}) that expands in a background thermal bath can
be written as [see eq. (\ref{18})]
\begin{eqnarray}
&& H^2_c(\phi_c) + H_c(\phi_c) H''_c(\phi_c)
\left<\phi^2\right> =
\frac{8\pi G}{3} \left[ \frac{\dot\phi^2_c}{2} \right. \nonumber \\
&& +\left.
V(\phi_c) + \frac{V''(\phi_c)}{2!} \left<\phi^2\right> \right], \label{18'}
\end{eqnarray}
where, in the left hand of eq. (\ref{18'}),
we have expanded to sencond order in $\phi$ the expectation
value of $H^2(\varphi)$.
Furthermore, in the right hand
we have neglected the terms $\left<(\nabla \phi)^2\right>/(2a^2)$ and
$\left<\dot\phi^2\right>/2$ because they are very small with respect
to the other on cosmological scales.

It is easy to show that the friction force generates entropy.
It was calculated by Hosoya and Sakagami [for details in the
calculation see\cite{1}]
\begin{equation}\label{18'''}
\dot S(t) = \frac{\eta_{eff}}{T} \left[\dot\phi_c\right]^2,
\end{equation}
where $S(t)$ is the entropy density.

To summarize, equations (\ref{16}), (\ref{17}), (\ref{18}) and
(\ref{18'}) describe completely
the dynamics of the background field $\phi_c$ on the thermal bath taking
into account the temporal evolution of the temperature due
to the expansion of the universe and the self-interaction of the
inflaton field in an universe globally isotropic, homogeneous and spatially
flat on cosmological scales which is described by a FRW metric.

\section{An example}

If we assume an $\phi_c \rightarrow -\phi_c$ invariant potential (\ref{12}),
the Hubble parameter that describes the evolution of the universe
can be obtained from the Friedmann equation (\ref{18'})
\begin{equation}\label{19}
H_c(\phi_c) = \sqrt{\frac{\pi G}{3}}\lambda \  \phi^2_c +H_0, 
\end{equation}
where 
$H_0=2\sqrt{{\pi G\over 3}} {m^2 \over \lambda}$ 
is the Hubble parameter at the end of inflation.

In this case $V'_{eff}(\phi_c,T)$ can be written
as
\begin{equation}          \label{20}
V'_{eff}(\phi_c,T) ={\cal M}^2(T) \  \phi_c +
\frac{\lambda^2_{eff}(T)}{6} \  \phi^3_c,
\end{equation}
where the effectives mass and self-interaction parameters are given by
\begin{eqnarray}
{\cal M}^2(T) & = & m^2 + \frac{\lambda^2 T^2}{24}
\left(1- \frac{4 m}{\pi^2 T}\right)
+ \frac{3. 2^6}{\pi} H_0 T, \label{20'}\\
\lambda^2_{eff}(T) & = & \lambda^2
+ 2^7 \sqrt{\frac{3^3 G}{\pi}}\lambda T. \label{21}
\end{eqnarray}
The third term in eq. (\ref{20'}) is related to the cosmological constant
$H_0 = \sqrt{{8\pi G \over 3}\Lambda}$. This means that the cosmological constant
should be $\Lambda = {m^4 \over 2 \lambda^2} \ll G^{-2}$. Furthermore, the
second term in (\ref{21}) is due to the thermal
effects in the self-interaction
of the inflaton field.

The temporal evolution of the inflaton field
goes as $t^{-1}$
\begin{equation}\label{22}
\phi_c(t) = \frac{\sqrt{6}}{\lambda t}.
\end{equation}

Furthermore,
in this case we obtain that the asymptotic value for the squared
field fluctuations $\left.\left< \phi^2(\vec x,t)\right>\right|_{t\gg t_0}$,
is
\begin{equation}\label{222}
\left.\left< \phi^2(\vec x,t)\right>\right|_{t\gg t_0}
\simeq \frac{T^2}{12} + \frac{3. 2^7}{\pi} H_0 T,
\end{equation}
where the last term is due to the interaction of the inflaton field
with the thermal bath in an expanding universe and the first one takes
into account this interaction without consider the expansion of the
universe. To obtain the temporal evolution of the temperature we must
to solve the eq. (\ref{14}). The exact solution is very difficult to
obtain, but the numerical calculation with $\lambda = 10^{-7} $ and
$m = 10^{-6} \  G^{-1/2}$ is shown in the figure (\ref{f151}). With the
parameter values here used we obtain $H_0 = 2.05 \  10^{-5} \  G^{-1/2}$
and $\Lambda = 5 \  10^{-11} \  G^{-2} $.
The effective friction parameter is many orders of magnitude bigger than
$H_c$ during all the inflationary stage, so that the thermodynamical
equilibrium is guaranteed. 
Furthermore, $\eta_{eff}$ in eq. (\ref{17}) approaches asymptotically
to $\eta$ during inflation, due to the last contribution in eq. (\ref{17})
becomes negligible for late times.

Figure (\ref{f151}) shows the evolution of the temperature $T(x)$, where
$x={\rm log}_{10}(t)$.
Note that the temperature increases with time until
$t\simeq 10^{4.1} \  G^{1/2}$, and it is sufficiently large to be possible
baryogenesis. After it, the temperature decreases slowly to
remain almost constant always up to $T\simeq 5 \  10^{-6} \  G^{-1/2}$.
Hence, at the
end of inflation the temperature remains with values bigger than the inflaton
mass ($T(x)/m > 5$), so that the high temperature limit used in the
calculation is always valid.
Furthermore, as can be shown the change of entropy (\ref{18'''}) is
always positive [see figure (\ref{f152})].
In the example here studied $\dot S$ reach its maximum value for times closed to
$t \simeq 10^{4.01} \  G^{1/2}$ to decrease monotonically
at the end of inflation.

At the end of inflation the fluctuations (\ref{222}) are
dominated by the last term: $\left.\left<\phi^2 \right>\right|_{e} \simeq
{3. 2^7 \over \pi} H_0 T \simeq 1.254 \  10^{-8} \  G^{-1}$ .
They are shown in the figure (\ref{f153}) and we can see that remains
almost constant at the end of inflation. The important fact here is
that thermal fluctuations are more important than quantum
fluctuations $\left<\phi^2\right>_{quant} \simeq H^2_c/(4\pi)$
in spite of the fact that $H_c < T$
during the last stages of inflation
in which the high temperature limit is valid.

On the other hand, the number of e-folds $N=\int^{t}_{t_0} H_c(t) dt$
is sufficiently large to assure the solution to the horizon/flatness
problems: $N[t>10^{6.5} \  G^{1/2}] > 60$.

\section{Final Comments}

In this paper we have used a consistent formalism to
find the effective equation of motion for the thermodynamical average
of the quantum scalar field $\varphi$. This equation agree with the
obtained by Hosoya and Sakagami, who obtained an additional term justified
by the change of temperature during inflation ($\dot T \neq 0$).
Such that a variation of temperature introduces a new term in the friction
parameter. However, as inflation evolves the additional term in $\eta_{eff}$
becomes negligible and $\eta_{eff} \simeq \eta$ at the end of inflation.

Furthermore, it is very interesting to note that the effective mass
of the inflaton field now has two additional terms; one is product of
the interaction [the second one in eq. (\ref{20'})]
and the other is originated
by the cosmological constant and the thermal
effects [the third one in eq. (\ref{20'})]. 
On the other hand, the thermal effects are also responsible for
an additional term in the effective self-interaction parameter
$\lambda_{eff}$ [see eq. (\ref{21})].
Notice that $\lambda_{eff} \rightarrow 0$ as $\lambda \rightarrow 0$.

To summaryze, in this paper I demonstrated that
thermal effects originated by dissipation
play an important role during inflation
and do not must to be ignored in models for the early universe.
Dissipation is so important that there are no oscillations of the
inflaton field when it reach the minimum energetic configuration
and at the same time, provides a smooth transition to the radiation
era.

\vskip .2cm
\centerline{{\bf ACKNOWLEDGEMENTS}}
\vskip .1cm
\noindent
I would like to acknowledge CONICET (Argentina) for partial
support and Universidad Nacional de Mar del Plata
for financial support in the form of a research grant.\\

\begin{center}
\begin{figure}
\includegraphics[width=8cm, height=7cm]{c:/tex/gr12003.bmp}
\vspace{-0.9cm}
\noindent
\caption{\label{f151} Evolution of
temperature $T(x)$. The time is shown in scale $x={\rm log}_{10}(t)$.} 
\end{figure}
\end{center}

\begin{center}
\begin{figure}
\includegraphics[width=8cm, height=7cm]{c:/tex/gr22003.bmp}
\vspace{-0.9cm}
\noindent
\caption{\label{f152} Evolution of
change of entropy $\dot S(x)$.
The time is shown in scale $x={\rm log}_{10}(t)$.}
\end{figure}
\end{center}

\begin{center}
\begin{figure}
\includegraphics[width=8cm, height=7cm]{c:/tex/gr32003.bmp}
\vspace{-0.9cm}
\noindent
\caption{\label{f153} Evolution of
fluctuations $\left<\phi^2\right>^{1/2}$.
The time is shown in scale $x={\rm log}_{10}(t)$.}
\end{figure}
\end{center}
\end{document}